\documentclass[letterpaper,prc,twocolumn,nofootinbib]{revtex4}

\usepackage{graphicx}
\usepackage{amsmath}
\usepackage{amssymb}
\usepackage[sort&compress]{natbib}
\usepackage{ifthen}
\usepackage{hyperref} %create internal and external hyperlinks
\usepackage{color}
\usepackage{horowitzarXiv}

\newcommand{\TITLE}{Measuring the Gluon Density in e + A Collisions:\\KLN CGC, DGLAP Glauber, or Neither?}

\definecolor{darkgreen}{rgb}{0,.7,0}
\definecolor{linkblue}{rgb}{0.,0.,0.9333}

\newcommand{\jpsi}{J/\psi}
\newcommand{\jpsit}{$\jpsi$ }

\hypersetup{
    pdffitwindow=true,      % page fit to window when opened
    pdftitle={\TITLE},    % title
    pdfauthor={W. A. Horowitz},     % author
    pdfnewwindow=true,      % links to different pdfs in new window
    bookmarksnumbered=true,
    colorlinks=true,       % false: boxed links; true: colored links
    linkcolor=red,          % color of internal links
    citecolor=darkgreen,        % color of links to bibliography
    filecolor=magenta,      % color of file links
    urlcolor=linkblue,           % color of external links
    menucolor=blue,
}

\begin{document}

\title{\TITLE}

\date{\today}

\author{W.\ A.\ Horowitz}
\email{wa.horowitz@uct.ac.za}
\affiliation{Department of Physics, University of Cape Town,\\Private Bag X3, Rondebosch 7701, South Africa}
\affiliation{Department of Physics, The Ohio State University,\\191 West Woodruff Avenue, Columbus, OH 43210, USA}

\begin{abstract}We predict readily experimentally measurable differences in the diffractive cross section in the coherent exclusive photoproduction of \jpsit mesons in e + A collisions at eRHIC and LHeC energies for nuclear gluon distributions assumed to 1) evolve in $x$ with DGLAP dynamics and have a spatial distribution proportional to the Glauber nuclear thickness function and 2) evolve in $x$ and $b$ according to the KLN prescription of CGC dynamics.  We find that CGC physics predicts that the nuclear gluon density widens significantly as a function of $x$ yielding diffractive peaks and minima that evolve dramatically with $x$; on the other hand the DGLAP Glauber distribution yields peaks and minima constant in $x$.  We also find that the dipole cross section at the level of two gluon exchange within the KLN parameterization of the CGC satisfies the black disk limit whereas this limit is violated when DGLAP evolution is used; the normalization of the diffractive cross section grows more slowly in $x$ by several orders of magnitude when using the KLN parameterization as compared to the result when employing DGLAP evolution.
\end{abstract}

\pacs{13.60.Hb, 24.85.+p}
\keywords{Small-x, DGLAP, Diffractive Vector Meson Production, CGC}

\maketitle

\section{Introduction}
Perturbative quantum chromodynamics (pQCD) predicts a nontrivial expansion in the size of the nuclear wavefunction at small $x$ due to the perturbative power law tails of the gluon distribution near the edge of the nucleus \cite{Froissart:1961ux,Gribov:1984tu,Iancu:2002xk,JalilianMarian:2005jf}.  Similarly, in order to not violate unitarity, the enormous growth in the gluon parton distribution function as $x$ becomes small found via na\"ive application of DGLAP evolution (see \cite{Nakamura:2010zzi} and references therein) must be tamed by perturbatively-calculable saturation effects \cite{Gribov:1984tu,Iancu:2002xk}.  However it is not yet clear from a theoretical standpoint at what values of $x$ these nontrivial changes in the dominant dynamics occur \cite{Iancu:2002xk}.   Additionally a quantitative theoretical understanding of experimental heavy ion data requires a quantitative understanding of the initial geometry of a heavy ion collision.  Certainly observables such as the azimuthal anisotropy of particles \cite{Hirano:2005xf,Luzum:2008cw,Jia:2010ee} is correlated with the anisotropy of the initial geometry; surprisingly the event-by-event fluctuations in the initial geometry also strongly affect these observables \cite{Schenke:2010rr,Jia:2011pi}.  In particular the viscosity to entropy ratio ($\eta/s$) of the quark-gluon plasma (QGP) found by comparing hydrodynamics simulations to heavy ion collision data is directly related to the eccentricity of the initial thermal quark-gluon plamsa distribution that is evolved hydrodynamically.  Currently the uncertainty in the initial thermal distribution due to the uncertainty in the importance of saturation effects in the initial nuclear profiles is large enough that it is not clear whether the physics of the QGP is better described by leading order weakly-coupled perturbative quantum chromodynamics (LO pQCD) or by LO strongly-coupled anti-de-Sitter/conformal field theory (AdS/CFT) methods \cite{Luzum:2008cw}.  An experimental measurement of the spatial gluon distribution in a highly boosted nucleus, and hence the relevant physics in this kinematic range, would thus be a very interesting and important contribution to our understanding of QCD.  

%%%%%%%%%%%%%%%%%%%%%%%%%%%%%%%%%%%%%%%%%%%%%%%%%%%%%%%%%%%%%%%%%%%%%%%%%%%%
\begin{figure}[!htb]
\centering
\includegraphics[width=3in]{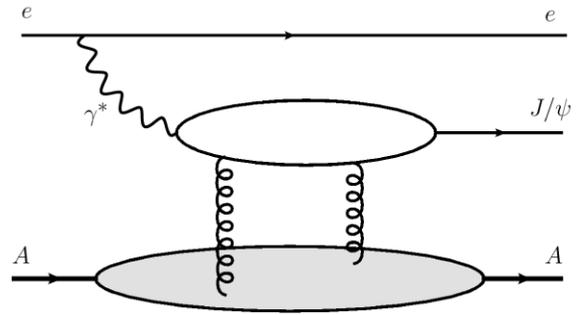}
\caption{\label{EVMPHU:handbag}Leading order Feynman diagram for the exclusive vector meson production of a \jpsit meson.}
\end{figure}
%%%%%%%%%%%%%%%%%%%%%%%%%%%%%%%%%%%%%%%%%%%%%%%%%%%%%%%%%%%%%%%%%%%%%%%%%%%%

%%%%%%%%%%%%%%%%%%%%%%%%%%%%%%%%%%%%%%%%%%%%%%%%%%%%%%%%%%%%%%%%%%%%%%%%%%%%
\begin{figure*}[!hbtp]
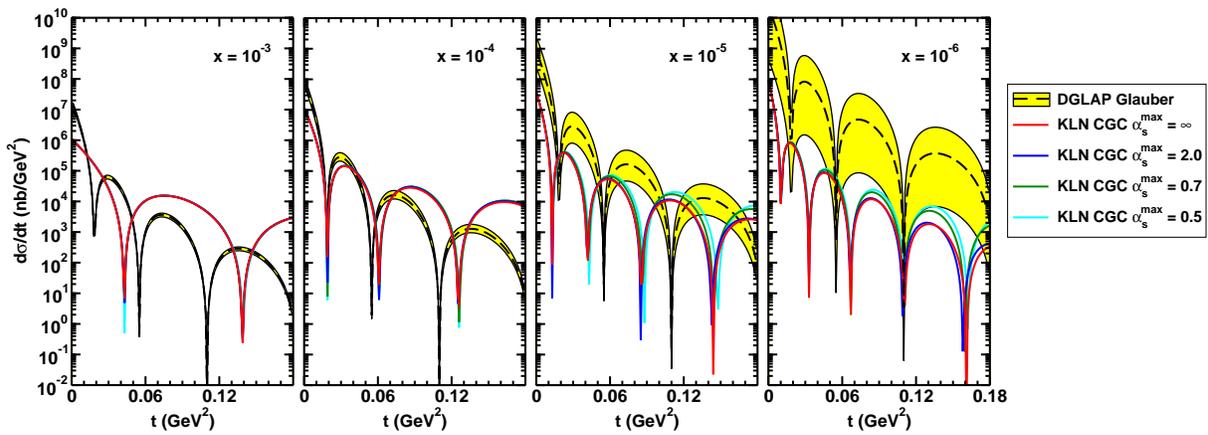

\centering
$\begin{array}{cccc}
\includegraphics[width=1.521in]{dsigdtx3b} & \hspace{-.07in}
\includegraphics[width=1.202in]{dsigdtx4b} & \hspace{-.07in}
\includegraphics[width=1.202in]{dsigdtx5b} & \hspace{-.07in}
\includegraphics[width=2.33in]{dsigdtx6d}
\end{array}$
\caption{\label{EVMPHU:dsigdtxdep}The diffractive cross section $d\sigma/dt$ from Eq.~\ref{EVMPHU:dsigdt} for 1) the DGLAP evolved dipole cross section, Eq.~\ref{EVMPHU:DGLAP}, with gluons spatially distributed according to the Glauber thickness function of the Woods-Saxon distribution and 2) for the dipole cross section from the KLN model of the CGC, Eq.~\ref{EVMPHU:CGCdipole}, for $x \, = \, 10^{-3},\,10^{-4},\,10^{-5},$ and $10^{-6}$.  The various $\alpha_s^{max}$ values shown explore some of the systematic theoretical uncertainty for the KLN CGC calculation, which is clearly much smaller than the difference between the results from DGLAP evolution and those from saturation physics. The yellow band represents the 1-$\sigma$ uncertainty (and the dashed black curve the central value) in the DGLAP Glauber results due to the uncertainty in the extracted LO MSTW gluon PDF \cite{Martin:2009iq}.}
\end{figure*}
%%%%%%%%%%%%%%%%%%%%%%%%%%%%%%%%%%%%%%%%%%%%%%%%%%%%%%%%%%%%%%%%%%%%%%%%%%%%

Exclusive vector meson production (EVMP) in e + A collisions has been proposed as a channel for just such a measurement \cite{Munier:2001nr,Caldwell:2010zza}.  In this Letter we will focus on the production of heavy vector mesons, in particular \jpsit mesons.  To leading order, EVMP of a \jpsit meson occurs in an e + A collision when a photon emitted by the electron splits into a $c$-$\bar{c}$ pair which communicates with the gluon density in the highly boosted nucleus via a two gluon exchange and subsequently forms a \jpsit meson and nothing else (we will be interested here in coherent EVMP, in which case the nucleus remains intact); see \fig{EVMPHU:handbag} for a visualization of the process.  It is precisely this two gluon exchange which yields a diffractive measurement of the gluon density in a nucleus; see \fig{EVMPHU:dsigdtxdep}.  

Previous work \cite{Caldwell:2010zza} explored how modest changes in the Woods-Saxon distribution \cite{Woods:1954zz} of a nucleus might manifest themselves as changes in the diffractive peaks in EVMP if one assumes that the spatial distribution of gluons in a nucleus is proportional to the Glauber thickness function found from the Woods-Saxon distribution.  That these modest changes do result in a visually obvious modification of the diffraction pattern motivated our further study, in which we consider whether two very different physical pictures of the gluon distribution in a highly boosted nucleus can be experimentally distinguished via EVMP: in particular we wish to compare the diffraction patterns that emerge when the gluon distribution 1) has normalization dictated by DGLAP evolution and spatial distribution given by the Glauber thickness function and 2) is given by the KLN parameterization (see \cite{Hirano:2004en,Kuhlman:2006qp} and references therein) of the Color Glass Condensate (CGC) (see, e.g., \cite{Iancu:2002xk,JalilianMarian:2005jf} for a review).  We choose to investigate these two ans\"atze of the gluon distribution in nuclei as they have been the dominant models used in heavy ion physics calculations to estimate the uncertainty in the viscosity to entropy ratio of the QGP produced at RHIC due to the uncertainty of the currently poorly constrained initial conditions in heavy ion collisions \cite{Hirano:2005xf,Luzum:2008cw}.  

It is worth taking a moment to comment on some common---yet confusing---terminology in the EVMP field.  As mentioned above, to leading order the coherent production of a vector meson in an e + A collision involves a two-gluon exchange between the $q$-$\bar{q}$ pair and the nucleus.  If one assumes that all two-gluon exchanges occur independently, then one may exponentiate the single two-gluon exchange result.  Making this independence assumption is often referred to in the EVMP field as using ``saturation'' physics because the cross section is unitarized via the exponentiation process.  However this ``saturation'' \emph{does not refer to small-$x$ evolution effects in the gluon distribution}.  For instance in the ``IP-Sat'' \cite{Kowalski:2003hm} and ``b-Sat'' \cite{Kowalski:2006hc} models, where ``Sat'' is short for saturation, the $x$ evolution of the gluon PDF is effected through the use of the DGLAP equations.  On the other hand, the ``b-CGC'' model \cite{Kowalski:2006hc} incorporates both the exponentiation of the two-gluon exchange \emph{and} the CGC physics of the saturation of the gluon PDF.  We note that, in principle, small-$x$ evolution effects and exponentiation effects in the dipole cross section should become appreciable simultaneously \cite{Mueller:1989st,Tuchin:2011dq}.  In order to (hopefully) make the presentation more clear, and to simplify some of the numerics, we will not exponentiate the two-gluon exchange; we will present results using only the leading order two-gluon exchange in which the gluon PDF is given either via DGLAP evolution or from the CGC.  Any subsequent reference to ``saturation'' in this paper will refer to the saturation of the gluon distribution function alone.

\section{Formalism}

Following \cite{Kowalski:2003hm,Caldwell:2010zza}, the diffractive production of a vector meson from a photon scattering off a target is
\begin{equation}
\label{EVMPHU:dsigdt}
\frac{d\sigma}{dt} = \frac{1}{16\pi}\left|\int d^2\boldsymbol{r}\int\frac{dz}{4\pi}\int d^2\boldsymbol{b} \, \langle V|\gamma\rangle_T \, e^{i\boldsymbol{b}\cdot\boldsymbol{\Delta}} \, \frac{d\sigma_{q\bar{q}}}{d^2\boldsymbol{b}}\right|^2,
\end{equation}
where $\langle V|\gamma\rangle_T$ is the overlap of the vector meson wavefunction and the transversely polarized virtual photon wavefunction---the contribution from the longitudinally polarized photon is zero as we are interested in $Q^2=0$ photoproduction---and we used the photon-meson overlap and Gauss-LC model for the \jpsit wavefunction from \cite{Kowalski:2003hm}\footnote{Note that the normalization of the \jpsit wavefunction in \cite{Kowalski:2003hm} is erroneously reported as a factor of 100 smaller than the correct value; one can readily see this by comparing with the normalization condition defined in \cite{Kowalski:2003hm} and with the results reported in \cite{Kowalski:2006hc}.  It is surprising that this error was not noted in \cite{Kowalski:2006hc}, in which the results found in \cite{Kowalski:2006hc} are explicitly compared to those in \cite{Kowalski:2003hm}.}, and $\boldsymbol{\Delta}^2 \, = \, -t$.  $d\sigma_{q\bar{q}}/d^2\boldsymbol{b}$ is the differential cross section for the interaction of the dipole with the target; its form depends on the physics assumptions we make for the nuclear gluon distribution, as we discuss in detail below.

\subsection{\texorpdfstring{DGLAP Evolution in $x$, Glauber Distribution of Gluons in $b$}{DGLAP Evolution in x, Glauber Distribution of Gluons in b}}

If we assume that the two gluon exchange from the dipole to the nucleus occurs within an individual nucleon then 
\begin{equation}
\label{EVMPHU:DGLAP}
\frac{d\sigma_{q\bar{q}}}{d^2\boldsymbol{b}} = \frac{\pi^2}{N_c} \, r^2 \, \alpha_s(\mu^2) \, xg(x,\,\mu^2) \, T(b),
\end{equation}
where $r$ is the size of the dipole, $\mu \, = \, \surd(\mu_0+C/r^2)$ is the relevant momentum scale for the dipole, $xg$ is the gluon distribution function, and
\begin{equation}
T(b) = \frac{1}{2\pi B_G}e^{-b^2/2B_G}
\end{equation}
is the assumed spatial distribution of gluons in a nucleon.  We use the MSTW parameterization of the gluon PDF \cite{Martin:2009iq}.  As described in \cite{Bartels:2002cj}, $\mu_0$ and $C$ are free parameters; as in \cite{Bartels:2002cj,Kowalski:2003hm,Lappi:2010dd}, we take $\mu_0$ = 1 GeV$^2$ and $C$ = 4.  From HERA data \cite{Chekanov:2002xi} the measured slope of %$d\sigma^{e+p\rightarrow e + p + \jpsi}/dt$
$d\sigma/dt$ yields $B_G \, \approx \, 4.25$ GeV$^{-2}$ \cite{Kowalski:2003hm}.  Then
\begin{equation}
\frac{d\sigma^{DGLAP}}{dt} = 4\pi \, \sigma_p^2 \, e^{-B_Gt} \, \left|\int db \, J_0(b\sqrt{t}) \, T_A(b) \right|^2,
\end{equation}
where $J_0$ is the usual Bessel function, 
\begin{align}
T_A(b) & \equiv \int dz \, \rho_A\left(\sqrt{b^2+z^2}\right) \\
\int d^2\boldsymbol{b} \, T_A(b) & = A,
\end{align}
is the usual thickness function, and $\rho_A$ is the density of the nucleus (here taken as the Woods-Saxon distribution of $^{197}$Au with the usual $R \, = \, 6.38$ fm and $a \, = \, 0.535$ fm \cite{Hahn:1956zz}) and
\begin{equation}
\sigma_p \equiv \frac{1}{4\pi}\int d^2\boldsymbol{r}\int dz \, \langle V|\gamma\rangle_T \, \frac{\pi^2}{N_c} \, r^2 \, \alpha_s(\mu^2) \, xg(x,\,\mu^2).
\end{equation}

%%%%%%%%%%%%%%%%%%%%%%%%%%%%%%%%%%%%%%%%%%%%%%%%%%%%%%%%%%%%%%%%%%%%%%%%%%%%
\begin{figure}[!htb]
\centering
\includegraphics[width=3in]{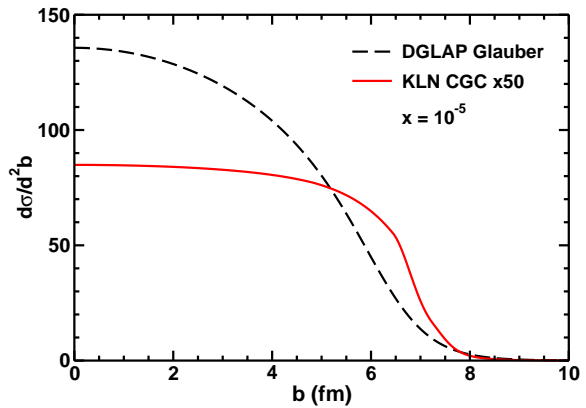}
\caption{\label{EVMPHU:density}The two gluon dipole cross section, $d\sigma_{q\bar{q}}/d^2\boldsymbol{b}$ from Eq.~\ref{EVMPHU:DGLAP} for the DGLAP Glauber calculation and Eq.~\ref{EVMPHU:CGCdipole} for the KLN version of the CGC (multiplied by 50), which is proportional to the gluon density distribution in a nucleus probed by a $q$-$\bar{q}$ dipole of size $r \, = \, 1$ GeV$^{-1}$ at $x \, = \, 10^{-5}$.}
\end{figure}
%%%%%%%%%%%%%%%%%%%%%%%%%%%%%%%%%%%%%%%%%%%%%%%%%%%%%%%%%%%%%%%%%%%%%%%%%%%%

\subsection{\texorpdfstring{CGC Distribution of Gluons in $x$ and $b$}{CGC Distribution of Gluons in x and b}}

Alternatively we may view the nucleus as a whole and that the gluon distribution is found from the CGC.  In this case
\begin{equation}
\label{EVMPHU:CGCdipole}
\frac{d\sigma_{q\bar{q}}}{d^2\boldsymbol{b}} = \frac{\pi^2}{N_C} \, r^2 \, \alpha_s(\mu^2) \, xg_A(\mu^2,\,Q_s^2),
\end{equation}
where $xg_A$ is the integrated gluon distribution function related to the unintegrated gluon distribution (UGD) $\phi_A$ by
\begin{align}
xg_A(\mu^2,\,Q_s^2) & = \int d^2\boldsymbol{k} \, \phi_A(k^2,\,Q_s^2) \nonumber\\
& = \pi\,\int_0^{k_{max}^2\,=\,\mu^2}dk^2\,\phi_A(k^2,\,Q_s^2)
\end{align}
The $x$ and $b$ dependence of the two-gluon exchange dipole scattering formula, Eq.~\ref{EVMPHU:CGCdipole}, comes in implicitly through the $x$ and $b$ dependence of $Q_s^2$ \cite{Kuhlman:2006qp},
\begin{equation}
\label{EVMPHU:Qs}
Q_s^2 \equiv \frac{2\pi^2}{C_F} \, \alpha_s(Q_s^2) \, xg(x,\,Q_s^2) \, T_A(b),
\end{equation}
where $C_F \, \equiv \, (N_c^2-1)/2N_c$.

In principle one determines the UGD via the JIMWLK evolution equations or, in the large-$N_c$ limit, the BK evolution equations (see \cite{Iancu:2002xk,JalilianMarian:2005jf} and references therein).  However, instead of solving the full evolution equations many heavy ion physics calculations use instead the KLN prescription of the CGC (see, e.g., \cite{Hirano:2004en,Kuhlman:2006qp}), which attempts to capture the main feature of CGC physics; in particular, the KLN UGD becomes saturated at momenta on the scale of the saturation scale $Q_s$.  Because of its widespread use in heavy ion physics calculations and in order to simplify our own calculations we, too, will use the KLN UGD,
\begin{equation}
\label{EVMPHU:phiKLN}
\phi^{KLN}_A(k,\,Q_s^2) = \frac{\kappa \, C_F \, Q_s^2}{2\pi^3\,\alpha_s(Q_s^2)}\left\{ 
\begin{array}{ll}
\left(Q_s^2+\Lambda^2\right)^{-1}, & \,k^2 \, \le \, Q_s^2 \\ 
\left(k^2+\Lambda^2\right)^{-1}, & \,k^2 \, > Q_s^2,\end{array}\right.
\end{equation}
where $\kappa$ is an $O(1)$ parameter meant to represent higher order corrections to the UGD, and $\Lambda \, = \, 0.2$ GeV \cite{Kuhlman:2006qp}.  

%%%%%%%%%%%%%%%%%%%%%%%%%%%%%%%%%%%%%%%%%%%%%%%%%%%%%%%%%%%%%%%%%%%%%%%%%%%%
\begin{figure*}[!htb]
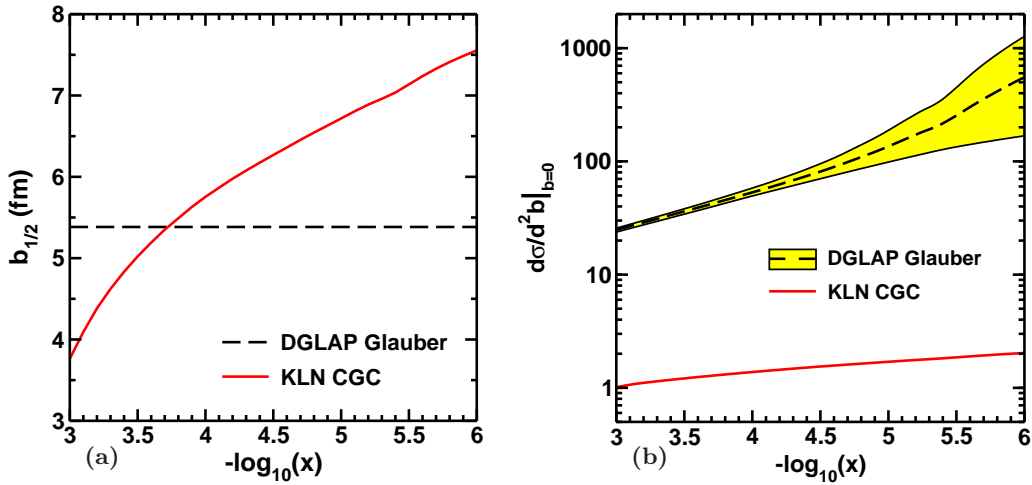

\centering
$\begin{array}{ccc}
\includegraphics[width=2.5in]{r12b} & &
\includegraphics[width=2.7in]{dsigdtb0b} \\[-.25in]
\hspace{-1.5in}\mathrm{\textbf{(a)}} & & \hspace{-1.3in}\mathrm{\textbf{(b)}}
\end{array}$
\caption{\label{EVMPHU:xdep}(a) $b_{1/2}$, the distance out from the center of the nucleus at which the dipole cross section is half its value at the center of the nucleus as a function of $x$. (b) $d\sigma_{q\bar{q}}/d^2\boldsymbol{b}|_{\boldsymbol{b}=0}$, the value of the dipole cross section evaluated at the center of the nucleus, as a function of $x$.  The yellow band represents the 1-$\sigma$ uncertainty (and the dashed black curve the central value) in the DGLAP Glauber results due to the uncertainty in the extracted LO MSTW gluon PDF \cite{Martin:2009iq}.  Both (a) and (b) are evaluated at dipole size $r \, = \, 1$ GeV$^{-1}$.}
\end{figure*}
%%%%%%%%%%%%%%%%%%%%%%%%%%%%%%%%%%%%%%%%%%%%%%%%%%%%%%%%%%%%%%%%%%%%%%%%%%%%

In principle $\kappa$ is set by comparing to known experimental observables such as the measured multiplicity at midrapidity at RHIC \cite{Back:2002uc,Adler:2004zn,Alver:2008ck} or LHC \cite{Aamodt:2010pb,Collaboration:2010cz} or to the diffractive cross sections for protons measured at HERA \cite{Chekanov:2002xi}.  However we found that the results from the leading order multiplicity formula \cite{Hirano:2004en} are linearly dependent on the cutoff taken for $\alpha_s$, $\alpha_s^{max}$.  The KLN UGD itself, though, is not nearly as sensitive to $\alpha_s^{max}$, so the multiplicity prescription does not provide a robust way of setting $\kappa$.  We note in passing that the centrality dependence of the particles produced via the leading order CGC multiplicity formula using the KLN UGD's also depends on $\alpha_s^{max}$.  Perhaps the use of the next-to-leading order results in the UGD \cite{ALbacete:2010ad} and/or the production formula \cite{Horowitz:2010yg} will mitigate this dependence enough to make reasonable comparisons of CGC multiplicity to current data.  Currently, though, there does not appear to be any quantitative estimate of the size of the dependence of the predicted CGC multiplicity as a function of centrality on $\alpha_s^{max}$.  $\kappa$ also cannot be set by comparing to the proton diffractive cross section as the currently available data does not probe regions of $x$ small enough such that $Q_s^2$ is a perturbative scale (at least when using the LO MSTW PDFs).  In our calculations we will set $\kappa \, = \, 1$.

It is important to contrast the interaction of the dipole in the KLN CGC approach taken here, in which the $q$-$\bar{q}$ pair interacts with the entire nucleus, and the Glauber approach, in which the pair interacts with individual nucleons.  By interacting with individual nucleons the diffractive cross section for the DGLAP Glauber model picks up an extra exponential suppression in $t$ proportional to the square of the width of the nucleon, $B_G$.  

\section{Results}

In \fig{EVMPHU:density} we plot $d\sigma_{q\bar{q}}/d^2\boldsymbol{b}$, which is directly proportional to the gluon density probed by the heavy quark dipole, for Eq.~\ref{EVMPHU:DGLAP} and Eq.~\ref{EVMPHU:CGCdipole}, the DGLAP and KLN CGC distributions, respectively.  The saturation physics of the CGC has resulted in a wider and flatter gluon distribution than that from the Glauber treatment; the DGLAP growth of the small-$x$ gluon distribution---tamed by the saturation physics of the KLN CGC---leads to a significant, nearly two orders of magnitude, enhancement in the cross section at $x \, = \, 10^{-5}$ compared to that found using the KLN CGC gluon distribution.  It is worth noting that \fig{EVMPHU:density} shows that the KLN prescription for the CGC satisfies the black disk limit even at the level of two gluon exchange whereas application of LO DGLAP evolution leads to a violation of the black disk limit by an order of magnitude at $x \, = \, 10^{-5}$.  

We attempt to quantify the changes in both the nuclear gluonic width and density as a function of $x$ in \fig{EVMPHU:xdep}.  In \fig{EVMPHU:xdep} (a) we show the quantity $b_{1/2}$, which we define as the radius at which the dipole cross section reaches half its value at the origin:
\begin{equation}
\frac{1}{2}\left.\frac{d\sigma_{q\bar{q}}}{d^2\boldsymbol{b}}\right|_{b \, = \, 0} \equiv 
\left.\frac{d\sigma_{q\bar{q}}}{d^2\boldsymbol{b}}\right|_{b \, = \, b_{1/2}},
\end{equation}
for the DGLAP and KLN CGC dipole cross sections.  We note that even out to extremely small values of $x\,\sim\,10^{-13}$, $b_{1/2}$ from the KLN CGC continues to rise sublinearly with $\log(s)$; thus the implementation of the KLN CGC used here, with the MSTW gluon PDF, satisfies the Froissart bound \cite{Froissart:1961ux}.  Intriguingly this sublinear (as opposed to linear) growth in radius as a function of $\log s$ is a surprise compared to other CGC parameterizations \cite{GolecBiernat:2003ym}.  In \fig{EVMPHU:xdep} (b) we show the dependence of the dipole cross section at $b\,=\,0$ on $x$ for the DGLAP Glauber and KLN CGC models.  Note the enormous growth of the dipole cross section as $x$ decreases for the LO DGLAP-evolved gluonic density.  This unitarity-violating enhancement is clearly reduced tremendously with the saturation physics of the KLN CGC.  The yellow band in the figure represents the 1-$\sigma$ uncertainty in the LO MSTW gluon PDF; the dashed black curve represents the result using the central value of the LO gluon PDF \cite{Martin:2009iq}.  

In \fig{EVMPHU:dsigdtxdep} we show the LO diffractive cross section for the EVMP of a \jpsit in e + A collisions, Eq.\ \ref{EVMPHU:dsigdt}, at $x \, = \, 10^{-3}, \, 10^{-4}, \, 10^{-5},$ and $10^{-6}$ when the gluon density grows in $x$ and $b$ according to DGLAP and Glauber overlap or KLN CGC.  As before, the yellow band describes the 1-$\sigma$ uncertainty in the LO MSTW gluon PDF, with the dashed black curve representing the central value.  Several KLN CGC curves are plotted; they correspond to the results when the maximum cutoff for $\alpha_s$, $\alpha_s^{max}$, is varied from $\infty$ down to $0.5$.  Note that all previous figures in this paper used $\alpha_s^{max} \, = \, \infty$.  For the various $\alpha_s^{max}$ curves in \fig{EVMPHU:dsigdtxdep}, the maximum value of the running coupling was set to $\alpha_s^{max}$ in: the dipole cross section, Eq.\ \ref{EVMPHU:CGCdipole}; the determination of the saturation scale, Eq.\ \ref{EVMPHU:Qs}; and also in the KLN UGD, Eq.\ \ref{EVMPHU:phiKLN}.  While an interesting question, the influence of the uncertainty in the gluon PDF on the saturation scale is beyond the scope of this work.  Clearly the KLN CGC diffractive cross section is not particularly sensitive to the specific $\alpha_s^{max}$ chosen, which implies that higher order running coupling corrections to the result are small.  The increase in the radial size of the gluon distribution as a function of $x$ shown in \fig{EVMPHU:xdep} (a) for the KLN CGC model manifests itself as a decrease in the spacings of the diffractive minima, $\Delta t_\mathrm{minima} \, \sim \, 1/b_{1/2}$, as one expects from a Fourier transform; on the other hand the positions in $t$ of the maxima and minima of the diffractive cross section for the DGLAP Glauber dipole do not change as a function of $x$.  

The drastically faster increase in the gluon density from the DGLAP evolved PDF seen in \fig{EVMPHU:xdep} (b) results in a cross section that increases much faster as a function of $x$ than for the KLN CGC case.  As was shown in \cite{Lappi:2010dd}\footnote{Figure 8 in \cite{Caldwell:2010zza} also shows that the incoherent process quickly dominates the coherent one as a function of $t$, although we note that there was an error in the calculation of the figure and that the curves plotted do not correspond to the equations in the text of the paper.} the incoherent cross section, in which the nucleus breaks up, begins to dominate the total diffractive cross section by $t \, \sim \, 0.02$ GeV$^{-2}$.  It is likely that the $t$ dependence of the incoherent EVMP of the two models will be different, although we do not provide a quantitative estimate here: the decrease in cross section as a function of $t$ for the DGLAP Glauber model will be enhanced by $\exp(-B_G\,t)$ due to the assumption that the heavy quark dipole interacts with individual nucleons.  And in the case of coherent scattering shown in \fig{EVMPHU:xdep}, one can discern a stronger $t$ dependence in the DGLAP Glauber results due precisely to the extra $\exp(-B_G\,t)$ factor that results from treating the nucleus as a collection of individual nucleons.  More importantly, the much larger gluon density yields a particularly noticeable difference at $t \, = \, 0$, where possible nuclear breakup effects are negligible: the DGLAP Glauber case is an order of magnitude larger than the KLN CGC case at $x \, = \, 10^{-3}$ and is a full two orders of magnitude larger at $x \, = \, 10^{-6}$.   Even with the very large PDF uncertainties as $x$ decreases, there is a clear increase in the coherent diffractive cross section for the DGLAP Glauber dipole compared to the KLN CGC dipole.  Note that the enormous normalization differences seen in \fig{EVMPHU:xdep} (b) between the DGLAP Glauber and KLN CGC dipoles for the most likely dipole size of $r \, = \, 1$ GeV$^{-1}$ for the photon-vector meson overlap do not directly translate into as large normalization differences in $d\sigma/dt$ due to the integration over all dipole sizes, $r$.

\section{Conclusions and Discussion}
An enormous wealth of information on the gluonic structure of highly relativistic nuclei can be found using exclusive vector meson production.  In particular we investigated the experimental signatures of the coherent scattering of a $c\bar{c}$ dipole onto a nucleus that results in an intact nucleus and a \jpsit meson in e + A collisions at eRHIC and LHeC energies.  We found that the diffractive cross section will readily experimentally differentiate between the two common initial highly boosted nucleus prescriptions used in heavy ion physics phenomenology: 1) the gluon density is found using DGLAP evolution and its spatial distribution is assumed to be proportional to the at-rest Glauber nuclear thickness function and 2) the gluon density and distribution is given by the KLN parameterization of the CGC.  In particular there is the exciting possibility of literally watching a nucleus grow with center of mass energy as the positions in $t$ of the minima and maxima in the diffractive cross section for the saturation physics calculation depend quite strongly on $\log(x)$.  On the other hand the DGLAP Glauber model yields a nucleus of constant size as a function of $x$; the positions in $t$ of the diffractive minima and maxima do not change as a function of $x$.  At the same time one is determining the width of a nucleus in e + A collisions, one will also measure the $x$ dependence of the normalization of $d\sigma/dt$.  Due to the explosion of small-$x$ gluons the DGLAP Glauber approach yields a normalization that rapidly increases as a function of $x$; additionally the $t$ dependence of the DGLAP Glauber $d\sigma/dt$ is also quite strong as it is proportional to $\exp(-B_G \, t)$ due to the assumption that the $q$-$\bar{q}$ dipole interacts with individual nucleons.  Conversely the KLN CGC dipole description does not have a strong $x$ dependence in its normalization due to its inclusion of saturation effects; similarly, the interaction of the dipole with the whole nuclear gluonic wavefunction yields a weaker $t$ dependence than is displayed by the DGLAP Glauber results.  

It is clear that, at the very least, the striking difference between the $x$ dependence of the peaks and minima from the DGLAP Glauber model and the KLN CGC model are robust: these differences will persist should we use even more sophisticated models of these two physical pictures; the $x$ dependence of the peaks and minima will persist should we attempt to approximate multiple scattering within the nucleus by exponentiating the dipole cross section, should we use a less approximate CGC calculation such as is found in \cite{ALbacete:2010ad}, or should we examine the results from other vector mesons such as the $\phi$ or $\rho$.  We regrettably leave the quantification of the diffractive cross section for these more sophisticated physical models and additional vector mesons for future work.  Exponentiating the two-gluon exchange cross section will reduce the enormous growth in the diffractive cross section in the DGLAP Glauber picture compared to the CGC case; we suspect this reduction will not be too large, although we also leave the quantification of this reduction to future work.

\subsection*{Acknowledgments}

The author wishes to thank Elke Aschenauer, Markus Diehl, Yuri Kovchegov, Henri Kowalski, Thomas Lappi, Cyrille Marquet, and Thomas Ullrich for invaluable discussions and the Institute for Nuclear Theory at the University of Washington for its hospitality and support.  The author wishes to especially thank Yuri Kovchegov for reading and commenting on the manuscript.  This research was sponsored in part by the U.S.\ Department of Energy under Grant No.\ DESC0004286.

\def\eprinttmppp@#1arXiv:@{#1}
\providecommand{\arxivlink[1]}{\href{http://arxiv.org/abs/#1}{arXiv:#1}}
\def\eprinttmp@#1arXiv:#2 [#3]#4@{\ifthenelse{\equal{#3}{x}}{\ifthenelse{
\equal{#1}{}}{\arxivlink{\eprinttmppp@#2@}}{\arxivlink{#1}}}{\arxivlink{#2}
  [#3]}}
\providecommand{\eprintlink}[1]{\eprinttmp@#1arXiv: [x]@}
\renewcommand{\eprint}[1]{\eprintlink{#1}}
\providecommand{\adsurl}[1]{\href{#1}{ADS}}
\renewcommand{\bibinfo}[2]{\ifthenelse{\equal{#1}{isbn}}{\href{http://cosmolog%
ist.info/ISBN/#2}{#2}}{#2}}

\end{document}